\documentclass[twocolappendix]{emulateapj}
\usepackage{amsmath}
\usepackage{subfigure}
\usepackage{graphicx}

\newcommand{\pd}[2]{ \frac{\partial #1}{\partial #2} }

\newcommand{\vsp}[1]{ \mathbf #1 }
\newcommand{\alfven}{Alfv{\'e}n }

\newcommand{\figext}[1]{#1.pdf}

\begin{document}

\title{Inverse cascade of non-helical magnetic turbulence in a relativistic fluid}
\shorttitle{Inverse cascade of relativistic MHD}

\author{Jonathan Zrake}
\shortauthors{J. Zrake}

\affil{Kavli Institute for Particle Astrophysics and Cosmology, Stanford
  University, Menlo Park, CA 94025, USA}


\begin{abstract}

The free decay of non-helical relativistic magnetohydrodynamic turbulence is
studied numerically, and found to exhibit cascading of magnetic energy toward
large scales. Evolution of the magnetic energy spectrum $P_M(k,t)$ is
self-similar in time and well modeled by a broken power law with sub-inertial
and inertial range indices very close to $7/2$ and $-2$ respectively. The
magnetic coherence scale is found to grow in time as $t^{2/5}$, much too slow to
account for optical polarization of gamma-ray burst afterglow emission if
magnetic energy is to be supplied only at microphysical length scales. No bursty
or explosive energy loss is observed in relativistic MHD turbulence having
modest magnetization, which constrains magnetic reconnection models for rapid
time variability of GRB prompt emission, blazars and the Crab nebula.



\end{abstract}

\maketitle



\section{Introduction}\label{sec:intro}
Freely decaying magnetohydrodynamic (MHD) turbulence is a phenomenon of
fundamental importance within the theory of magnetized fluids. That its
operation may include the cascading of energy toward larger scales bears
far-reaching implications in cosmology and high-energy astrophysics. For
example, the strength and coherence scale of the present-day galactic magnetic
field could be explained by inverse cascading from extremely small scale fields
seeded by phase transitions in the early universe \citep{Field2000,
  Tevzadze2012}. Inverse cascading of magnetic energy could also explain recent
measurements of strong optical polarization in gamma-ray burst (GRB) afterglows
\citep{Uehara2012, Mundell2013}, where magnetic energy production is believed to
operate only at very small scales.

Turbulent inverse cascades are associated with the accumulation of energy at
wavelengths longer than the turbulence integral scale. They entail the
self-organization of turbulent structures, wherein order emerges from chaotic
initial conditions. A familiar example is that of two-dimensional hydrodynamic
turbulence, where inverse cascading of kinetic energy is a consequence of global
enstrophy conservation. Inverse cascades are qualitatively distinct from direct
cascades in that they shift energy away from, rather than toward the dissipation
scale. In general, turbulent energy flux moves in both directions. But in
three-dimensional hydrodynamic turbulence, modes above the integral scale are
damped by instabilities faster than they are pumped by motions in the inertial
range.




Since the work of \cite{Frisch1975} it has been well appreciated that MHD
turbulence may exhibit inverse cascading as a consequence of global magnetic
helicity conservation. But the literature to date is still conflicted on
whether helicity is a necessary condition for inverse cascading to occur. It
was shown by \cite{Olesen1997} and \cite{Shiromizu1998} that inverse cascading
could be expected even for non-helical configurations, as a consequence of
rescaling symmetries native to the Navier-Stokes equations. But no inverse
cascading was seen in numerical studies based on EDQNM theory \citep{Son1999}
or direct numerical simulations with relatively low resolution
\citep{Christensson2001, Banerjee2004}. Given that mechanisms for helicity
production in the early universe are uncertain, and completely absent from
regions of GRB afterglow emission, it is crucial to understand the operation of
freely decaying non-helical MHD turbulence.

\begin{figure*}\label{fig:images}
  \centering
  \includegraphics[width=7.2in]{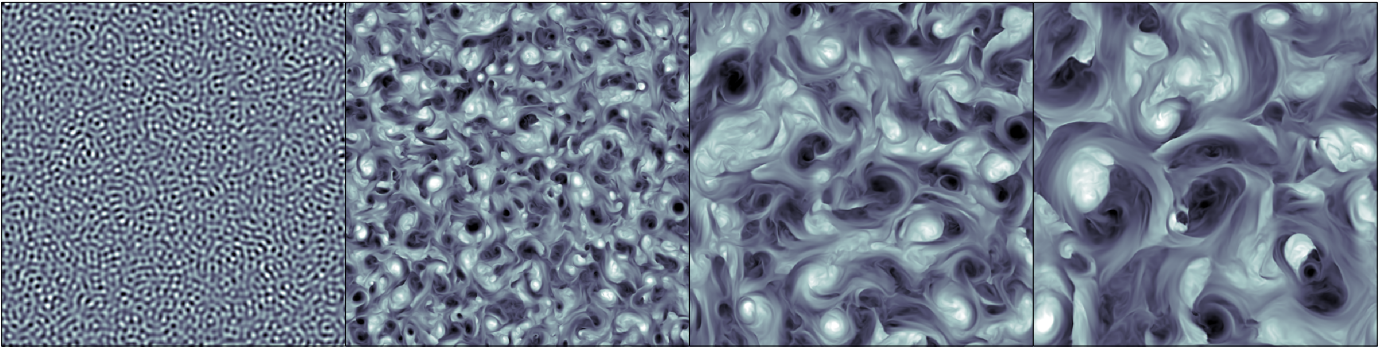}
  \caption{Two-dimensional slices of transverse magnetic field component showing
    the progression of magnetic field decay in a three-dimensional relativistic
    MHD turbulence. The left-most panel shows the initial condition, and then
    from left to right the solution is shown at 4, 32, and 128 initial \alfven
    crossing times of the simulation domain.}
\end{figure*}

In this Letter we establish that helicity is not a necessary condition for
inverse cascading in relativistic MHD turbulence. The intended domains of
applicability are the evolution of primordial magnetic fields, and those thought
to be responsible for the synchrotron emission of GRB afterglows. Given that
neither is free of relativistic complications, our results are based on
numerical solutions of the relativistic MHD equations. We adopt the initial
value problem $P_M(k,0) \propto \delta(k - k_0)$, where $k_0^{-1}$ is much
smaller than the simulation domain ($P_M(k,t)$ is defined so that the
electromagnetic energy density $E_M(t) = \int P_M(k,t)~dk$). This choice is
permits the system to evolve toward a universal energy spectrum, allowing the
sub-inertial and inertial range indices to be measured instead of imposed.

Numerical simulations exhibiting inverse cascades in non-helical,
non-relativistic MHD turbulence were reported by \cite{Brandenburg2014}
concurrently with the preparation of this work. Our treatment goes farther by
including relativistic effects, and by proposing a self-similar ansatz for the
evolution of $P_M(k,t)$ which agrees very closely with the simulation
results. We have studied freely decaying MHD turbulence, whereas
\cite{Brandenburg2014} assumed continuous magnetic energy injection at small
scales. Despite these differences, both studies support the existence of inverse
magnetic energy transfer in non-helical MHD turbulence.  The case of
relativistic MHD turbulence driven continuously at large scales as been treated
previously \citep{Zrake2011, Zrake2013}. Our numerical setup is described in
Section \ref{sec:setup}. Simulation results and our self-similar ansatz are
given in Section \ref{sec:results}. In Section \ref{sec:discussion-picture} we
suggest a phenomenological picture that accounts for inverse cascading of MHD
turbulence. We also draw comparisons with previous numerical and analytical work
in Section \ref{sec:discussion-compare}, and in Section \ref{sec:discussion-ivp}
examine the generality of the initial value problem chosen for this
study. Finally, in Section \ref{sec:discussion-observational} we discuss the
implications of our findings to the physics of GRB prompt and afterglow
emission.

\section{Numerical set-up}\label{sec:setup}
The scenario investigated here is described as follows. Consider a perfectly
conducting fluid whose rest mass, thermal, and magnetic energy densities are
mutually comparable. Assume that the magnetic field has periodicity scale $L$,
is out of equilibrium such that $\mathbf{J} \times \mathbf{B} \ne 0$, is
non-helical, and has an energy spectrum $P_M(k,0)$ that is peaked at the scale
$k_0 \gg 2\pi/L$. Time-dependent solutions of the relativistic MHD equations
\begin{subequations}\label{eqn:rmhd-system}
  \begin{align} \nabla_\nu \rho
    u^\mu &= 0 \\ \nabla_\nu T^{\mu \nu} &= 0 \\
    \pd{\vsp B}{t} &= \vsp \nabla
  \times (\vsp v \times \vsp B)
  \end{align}
\end{subequations}
are obtained using the \texttt{Mara} code \citep{Zrake2011} run on a
three-dimensional computational mesh with $512$ grid points along each axis.  In
Equation \ref{eqn:rmhd-system}, $T^{\mu \nu}$ is the stress-energy tensor
including both hydrodynamic and electromagnetic contribution, $u^\mu$ is the
fluid four-velocity, and $\rho$ is mass density. The magnetic field is initially
divergenceless and Gaussian-random with a power spectrum that is narrowly peaked
around the wavenumber $k_0 = 50 k_1$, where $k_1 = 2\pi/L$. $P_M(k,t)$ is
normalized so that the plasma-$\beta$, the ratio of gas to magnetic pressure, is
initially $1$.

We define inverse cascading as the accumulation of energy in the sub-inertial
range modes (those above the turbulence integral scale), which is evident when
the magnetic energy spectrum $P_M(k,t)$ is an increasing function of time for
wavenumbers $k < k_t$ where $k_t$ is integral scale wavenumber at time $t$. Note
that migration of $k_t$ toward smaller values over time is not a sufficient
condition for inverse cascading; growth of the coherence scale also occurs in
so-called ``selective decay'', whereby energy is processed through a direct
cascade that drains energy in the small scales before the larger. Interestingly,
both processes have been suggested to involve leftward migration of $k_t$
depending upon time like $t^{-2/5}$ \citep{Olesen1997, Shiromizu1998, Son1999}.

\section{Results}\label{sec:results}
Figure \ref{fig:images} shows two dimensional slices of the out-of-page magnetic
field component taken at roughly logarithmic intervals throughout the
simulation. The left-most panel shows the initial Gaussian-random magnetic field
configuration. The second panel shows the solution after a single \alfven
crossing time of the simulation domain, during which the field has organized
itself into a collection of small magnetic islands having complex internal
structure. The third and fourth panels show those islands becoming larger in
scale, and less numerous. The color mapping has been stretched to the minimum
and maximum data values of each image, so only the field morphology is depicted
and not its average magnitude. Since the initial condition lacks magnetic energy
at large scales, the appearance of larger coherent magnetic field structures
cannot be selective decay, but can only be attributed to the inverse transfer of
magnetic energy from small to large scales.

\begin{figure}
  \centering
  \includegraphics[width=3.6in]{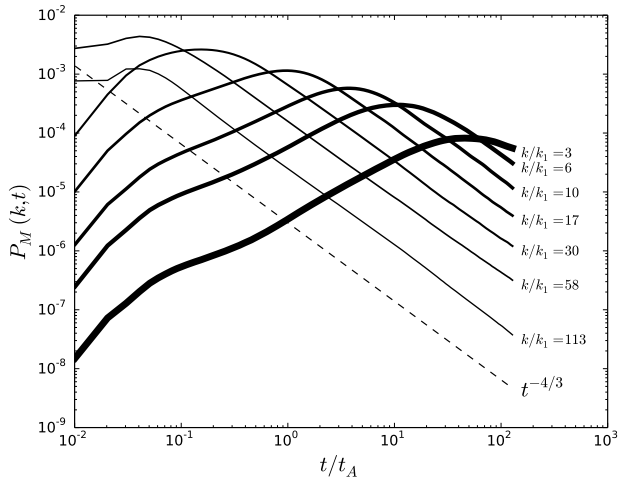}
  \caption{The temporal evolution of $P_M(k,t)$ at seven representative
   wavenumbers. Heavier ink denotes larger scales. The dashed line shows a power
   law with index $-4/3$}
  \label{fig:Pmkt1}
\end{figure}

Indeed, as shown in Figure \ref{fig:Pmkt1} the magnetic energy spectrum
$P_M(k,t)$ is an increasing function of time for small $k$ at early times. For
each wavenumber $k<k_0$, there is a turn-over time $\tau_{k}$ when
$\frac{\partial}{\partial t}P_M(k,t)$ switches sign. $\tau_k$ is thus the time
when coherent magnetic field structures of wavenumber $k$ are fully developed,
and captures the time required for the magnetic field to assemble itself at
length scale $k^{-1}$. At times $t > \tau_k$, the amplitude of wavenumber $k$
structures diminishes as a power law in time, $P_M(k,t) \propto t^{\delta}$
where $\delta$ is measured to be $-1.3 \pm 0.03$. The fiducial value of $-4/3$
will be adopted for simplicity.

\begin{figure}
  \centering
  \includegraphics[width=3.6in]{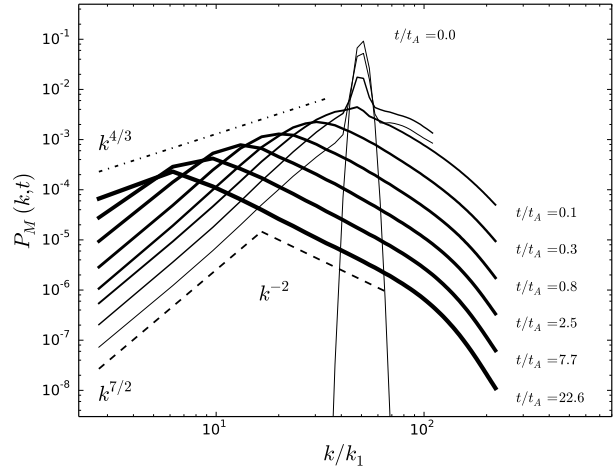}
  \caption{$P_M(k,t)$ shown at nine representative times, including $t=0$ and
  proceeding through $t=22.6 t_A$ with lines of increasing width. The dashed
  lines show power laws with indices $3.5$ and $-2$ for the scales larger and
  smaller than the injection scale $2\pi/k_0$ respectively. The dashed-dotted
  line shows $P_M(k,\tau_k) \propto k^{4/3}$.}
  \label{fig:Pmkt2}
\end{figure}

Figure \ref{fig:Pmkt2} shows $P_M(k,t)$ at several times throughout the
simulation. After a fraction of an \alfven time, the magnetic energy spectrum
relaxes to a form which is well described by a split power law
\begin{equation}
  P_M(k,t_A) \propto
  \begin{cases}
    \left(\frac{k}{k_0}\right)^{\alpha} & k < k_0 \\
    \left(\frac{k}{k_0}\right)^{\beta} & k \ge k_0
  \end{cases}.
  \label{eqn:split-plaw}
\end{equation}
where the sub-inertial and inertial range indices are measured to be $\alpha =
3.50 \pm 0.04$ and $\beta = -1.91 \pm 0.005$ respectively. The values $\alpha =
7/2$ and $\beta = -2$ will be adopted for simplicity. We note here that the
magnetic energy spectrum is found to be significantly steeper than $5/3$ as is
predicted in the Goldreich-Sridhar \citep{Goldreich1995} phenomenology. $5/3$
scaling has been verified numerically in strong \alfven wave turbulence as well
as isotropic MHD turbulence driven kinetically at large scales \cite[see
  e.g.][for a review]{Tobias2011}. However, it appears that isotropic,
\emph{freely decaying} MHD turbulence has a slope that is significantly steeper
than is predicted by the Goldreich-Sridhar theory.

\begin{figure}
  \centering
  \includegraphics[width=3.6in]{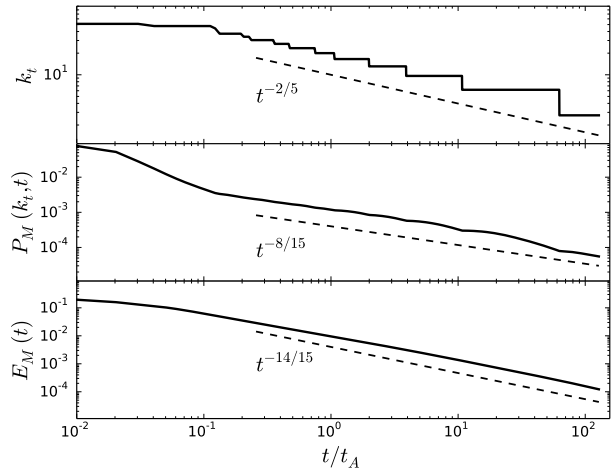}
  \caption{The upper panel shows the peak wavenumber $k_t$ as a function of
    time, alongside a power law of index $-2/5$ for comparison to analytic
    predictions. The steps are an artifact of the finite spectral resolution.
    The middle panel shows $P_M(k_t,t)$ as a function of $t$, and the power law
    with index $-8/15$ predicted by the empirical model. The lower panel shows
    the evolution of the average magnetic energy density $E_M(t)$, which
    deviates slightly from the reference slope of $-14/15$ due to the finite
    value of $L k_t$.}
\label{fig:kmt}
\end{figure}

As shown in the upper panel of Figure \ref{fig:kmt}, the break in the power
spectrum lies at $k_t \propto t^{\gamma}$ where $\gamma$ is consistent with the
value of $-2/5$ predicted by scaling arguments made in \cite{Shiromizu1998} and
\cite{Olesen1997}.
Throughout the simulation, the sub-inertial and inertial range indices remain
fixed, with the peak of magnetic energy moving down and to the left on the axes
of Figure \ref{fig:Pmkt2}. In other words, the evolution of the magnetic energy
spectrum is very nearly self-similar, being well-described by
\begin{equation}
  P_M(k,t) = s^{\gamma\beta + \delta} P_M(k s^{-\gamma},t_A)
\end{equation}
where $s = t/t_A$ and $\delta = -4/3$ is the power-law index for decay at all
wavenumbers larger than $k_t$, as shown in Figure \ref{fig:Pmkt1}. In this
empirical model the magnetic energy at each scale larger than $k_t^{-1}$ grows
proportionally to $t^{\gamma (\beta - \alpha) + \delta} = t^{13/15}$ and the
energy associated with peak magnetic structures, $P_M(k_t,t)$ diminishes as
$t^{\gamma \beta + \delta} = t^{-8/15}$. Those peaks trace out $P_M(k,\tau_k)
\propto k^{4/3}$ as shown in the dashed-dotted line of Figure \ref{fig:Pmkt2}.
In the limit of $L k_t \rightarrow \infty$ the total magnetic energy $E_M(t)
\propto t^{\gamma (\beta + 1) + \delta} = t^{-14/15}$ as shown in the lower
panel of \ref{fig:kmt}.

\section{Discussion}\label{sec:discussion}
\subsection{Comparison with other studies}\label{sec:discussion-compare}
Direct numerical simulation of freely decaying non-helical MHD turbulence have
been carried out by \cite{Christensson2001} and \cite{Banerjee2004} which report
selective decay and no inverse cascade. Nevertheless, it is possible that an
inverse cascade was present, but hidden beneath the sub-inertial part of the
imposed energy spectrum, for which indices of $2$ and $4$ were chosen by each
study respectively. It was observed here that the locus of peak spectral energy
$P_M(k,\tau_k) \propto k^{4/3}$, so additional scale separation might have been
required in those studies for an inverse cascade to become apparent.  Our
results are in general agreement with those of \cite{Brandenburg2014}, which are
based on direct numerical simulations of non-helical, non-relativistic MHD
turbulence done with very high resolution. That study reported a slightly
steeper slope of the sub-inertial range.

Inverse cascading of magnetic energy in the test-field limit was also reported
very recently by \cite{Berera2014}. This study found that passive vector fields
advected within fully developed, isotropic hydrodynamic turbulence attain
coherency over increasing sptial scales. This discovery offers an interesting
avenue to examining the generality of inverse cascading.

The inverse cascading observed in our study is not the result of residual
helicity in the initial data. Helicity conservation requires only that the
correlation scale $k_t^{-1}$ is larger than $k_M^{-1} = H_M / E_M$
\citep{Tevzadze2012}. But in our study $k_M^{-1}$ evolves from $1/1000$ of
the grid spacing up to roughly the grid spacing throughout the simulation. So
in fact the correlation scale $k_t^{-1}$ remains at least 1000 times
larger than the lower limit imposed by helicity conservation throughout the
simulation.

\subsection{Generality of the initial value problem}\label{sec:discussion-ivp}
We have found that inverse cascading of magnetic energy proceeds from the
initial value problem $P_M(k,0) \propto \delta(k - k_0)$. After a fraction of an
\alfven time, the spectrum relaxes toward the split power law in Equation
\ref{eqn:split-plaw}. Subsequent enhancement of magnetic energy at scales larger
than $k_t^{-1}$ occurs through self-similar evolution of the split power
solution. Thus inverse cascading must also occur for any initial value problem
$P_M(k,0)$ which first evolves toward the split power-law solution. Initial
value problems where $P_M(k,0)$ has non-compact support have also been
considered. \cite{Olesen1997} predict that inverse cascading from power-law
initial data occurs if and only if the sub-inertial range index $\alpha > -3$,
regardless of the magnetic helicity. Verifying this claim numerically will be
the topic of a future study.



\subsection{Phenomenological picture}\label{sec:discussion-picture}
We propose that inverse cascading manifests as ``unwinding'' and ``re-linking''
of the magnetic field lines. Unwinding refers to the field's preference for
configurations in which the tension force $\mathbf{B} \cdot \nabla \mathbf{B}$
is more uniformly distributed in space. Re-linking occurs where magnetic field
loops sourced by parallel line currents attract one another by their mutual
Lorentz force $\mathbf{J} \times \mathbf{B}$. This brings regions of opposing
magnetic flux into contact with one another creating X-point reconnection
sites. At those sites the two flux loops are joined into a single one shaped
like a peanut, which then tries to attain maximal average curvature by
deforming itself into a circle. This process has the distinct effect of
assembling coherent magnetic structures over progressively larger scales, and
is insensitive to the mutual linking between flux loops, i.e. the magnetic
helicity.

In this scenario inverse magnetic energy transfer cannot be avoided, but its
efficiency depends inversely upon the energy lost by the magnetic field during
and immediately after the reconnection event. Some portion of magnetic energy is
thermalized by non-ideal effects at the moment of reconnection, and another
portion is transferred to the bulk flow by accelerating fluid away from the
reconnection site. Maximal efficiency of the inverse cascade is attained when
the expansion of magnetic field loops is fully adiabatic, in which case the
magnetic energy density $E_M \propto \ell^{-2}$ where $\ell$ is the size of the
loop. Since the characteristic size of the loops scales as $\ell \propto
t^{2/5}$, the shallowest possible decay law allowed by this model is $E_M
\propto t^{-4/5}$. That is only marginally shallower than the decay of
$t^{-14/15}$ reported here, which is in turn considerably shallower than
$t^{-6/5}$ which is expected if all the energy is lost irreversibly to a direct
cascade as in Saffman's law \citep{Saffman1967}. The steeper $t^{-6/5}$ follows
from the assumption that a fixed fraction of magnetic energy at scale $\ell$ is
lost every \alfven time, which grows as $\sim \ell / v_A \propto t^{2/5}
E_M^{-1/2}$ due to the increasing coherence length and decreasing \alfven speed.

This picture of hierarchical merging of magnetic islands may be consistent with
the turbulent reconnection process studied in \cite{Lazarian1999} and
\cite{Eyink2011a, Eyink2013}. Turbulent reconnection predicts small energy
losses to direct heating, consistent with what has been reported here. Evidence
for low thermalization rates has also been found in recent kinetic simulations
of magnetic reconnection across a single current sheet in both relativistic
\citep{Sironi2014} and non-relativistic \citep{Dahlin2014} plasmas. Those
studies show that direct heating caused by parallel electric fields at
reconnection sites may be weaker than was previously believed. Instead, the
magnetic energy liberated by the change in field topology goes largely into
accelerating the bulk flow away from the reconnection site.





\subsection{Observational implications}\label{sec:discussion-observational}
Optical polarization recently detected in GRB afterglows \citep{Uehara2012,
  Mundell2013} requires the magnetic field to attain coherency over the emitting
region. If the magnetic field is incoherent immediately behind the shock, the
coherence scale would have to grow like $\lambda_t \propto t$ to account for the
polarized afterglows \citep{Gruzinov1999a}. Given that non-helical MHD
turbulence has been found to decay with $\lambda_t \propto t^{2/5}$, there is
simply no way for non-linear evolution of the magnetic field in the downstream
region to account for this polarization. If detections of polarized afterglows
are to continue, it would be compelling evidence that the magnetic field is
already coherent across the emitting region when it is produced at the shock
front. This, in turn would favor turbulent dynamo mechanisms for the magnetic
energy production \citep{Milosavljevic2007, Sironi2007, Goodman2008,
  Duffell2014} over those based on microphysical plasma instabilities
\citep{Gruzinov1999a, Spitkovsky2008, Keshet2009, Sironi2009, Sironi2013a}.

The nature of freely decaying relativistic MHD turbulence, as presented here,
also bears implications for the rapid variability of GRB prompt emission
\citep{Usov1994, Gehrels2009}, blazars \citep{Sikora2009, Harris2009,
  Harris2011, Hayashida2012, Bhatta2013}, and the Crab nebula \citep{Tavani2011,
  Abdo2011}. The explosive release of magnetic energy by spontaneous magnetic
reconnection has also been implicated as a possible mechanism for the flares in
GRB prompt emission \citep{Lyutikov2003, Narayan2009, Zhang2011, Zhang2014},
blazars \citep{Giannios2009a, Giannios2010, Nalewajko2011, Calafut2014,
  Marscher2014}, and the Crab nebula \citep{Clausen-Brown2012,
  Cerutti2012}. However, our simulations, carried out for modest magnetizations
with the plasma-$\beta$ initiated at $1$, did not show any indication of bursty
or explosive magnetic energy loss; the decay profile is smooth and fluid motions
remain slightly below the \alfven speed at all times. Such behavior is expected
given that reconnection in the turbulent fluid takes place over the range of
turbulent length-scales.

The absence of reconnection bursts implies that flaring events, if indeed they
come from turbulent reconnecting magnetic fields, can only originate in
magnetically dominated plasmas if at all. Simulations of freely decaying
relativistic, magnetically dominated MHD turbulence are currently being
pursued. It is anticipated that in the magnetically dominated case where the
\alfven speed approaches that of light, the turbulent bulk flow will also become
relativistic. If so, then models invoking relativistic turbulence
\citep[e.g.][]{Narayan2009} to explain high energy flaring events may remain
viable.

\acknowledgments
The author wishes to thank F. Fiuza, T. Abel, and W. East for inspiring
discussions. This research was supported in part by NASA and by the NSF through
grant AST-1009863. Resources supporting this work were provided by the NASA
High-End Computing (HEC) Program through the NASA Advanced Supercomputing (NAS)
Division at Ames Research Center.

\bibliographystyle{apj}


\end{document}